\documentclass[aps,epsfig,psfig,prl,twocolumn,superscriptaddress,showpacs,showkeys,email]{revtex4}
\usepackage{graphics}
\begin{document}

\title{\bf Nucleation phenomena in protein folding: The modulating role of 
protein sequence}

\author {RDM Travasso}
\email{rui@cii.fc.ul.pt}
\author{PFN Fa\'\i sca}
\email{patnev@cii.fc.ul.pt}
\affiliation{Centro de Fisica Teorica e
Computacional da Universidade de Lisboa, Av. Prof. Gama Pinto 2, 1649-003
Lisboa Codex, Portugal}
\author{MM Telo da Gama}
\email{margarid@cii.fc.ul.pt}
\affiliation{Centro de Fisica Teorica e
Computacional da Universidade de Lisboa, Av. Prof. Gama Pinto 2, 1649-003
Lisboa Codex, Portugal}

\pacs{\bf{87.15.Cc; 91.45.Ty}}
\keywords{\bf{lattice models, Monte Carlo simulation, folding
nucleus, kinetics }}

\begin{abstract}
{\bf For the vast majority of naturally occurring, small, single domain 
proteins folding is often described as a two-state process that lacks 
detectable intermediates. This observation has often been rationalized on 
the basis of a nucleation mechanism for protein folding whose basic premise 
is the idea that after completion of a specific set of contacts forming the 
so-called folding nucleus the native state is achieved promptly. Here we 
propose a methodology to identify folding nuclei in small lattice polymers 
and apply it to the study of protein molecules with chain length N=48. 
To investigate the extent to which protein topology is 
a robust determinant of the nucleation mechanism we compare the nucleation 
scenario of a native-centric model with that of a sequence specific model 
sharing the same native fold. To evaluate the impact of the sequence's finner
details in the nucleation mechanism we consider the folding of two non-
homologous sequences. We conclude that in a sequence-specific model the 
folding nucleus is, to some extent, formed by the most stable contacts in the 
protein and that the less stable linkages in the folding nucleus are solely 
determined by the fold's topology. We have also found that independently of 
protein sequence the folding nucleus performs the same `topological' 
function. This unifying feature of the nucleation mechanism results from the 
residues forming the folding nucleus being distributed along the protein 
chain in a similar and well-defined manner that is determined by the fold's 
topological features. }
\end{abstract}

\maketitle

\section{Introduction}

Proteins do not appear to fold by means of a unique mechanism and 
over the years several phenomenological models have been proposed for
protein folding ~\cite{1,2,3,4,5,6,7,8,9,Brog,guid}. The framework model, for
example, is based on the idea that the formation of the
hydrogen-bonded secondary structural elements precedes the formation
of tertiary structure~\cite{1,2}, and the diffusion-collision model
assumes that part of the protein folding process involves the
interaction of metastable regions of structure which, when in contact,
may provide additional stabilization~\cite{3}.\par  
Chymotrypsin inhibitor 2, a small, single domain, two-state folder with
64 residues, epitomizes the so-called nucleation-condensation
(NC) mechanism for protein folding. The latter was firstly 
investigated by Shakhnovich, in the context of Monte Carlo lattice
simulations~\cite{4,5}, and by Fersht through extensive protein
engineering studies~\cite{6} termed $\phi$-value analysis. 
The NC mechanism can be viewed as a modified version of the
nucleation-growth mechanism originally proposed by
Wetlaufer~\cite{7}. The basic premise of the NC model is  
the idea that once a specific set of contacts, named the folding
nucleus (FN), forms there is a concerted consolidation of 
secondary and tertiary interactions as the whole protein  rapidly 
collapses to the native fold. \par 
More recently, the topomer search model, which emphasizes native state's 
topology as a major determinant of protein folding rates has been 
proposed~\cite{9} and investigated in the context of off-lattice Langevin
simulations~\cite{10, 11}. While 
it seems well established that the native topology, as measured by the
contact order parameter~\cite{12}, and other related
quantities~\cite{13,14,15}, is a major determinant of two-state
protein folding kinetics, the question of understanding the relative
roles played by native structure~\cite{16} and protein
sequence~\cite{17} in determining the folding mechanism remains to be
elucidated (reviewed in~\cite{18}).\par            
In their seminal work~\cite{4}, Abkevich and coworkers have found
that native structure is a more robust determinant of the
folding mechanism than sequence 
for 36-mer lattice proteins. Indeed, the results of Monte Carlo
simulations reported by Abkevich and coworkers~\cite{4} suggest that
three non-homologous sequences sharing the same native fold also share
a common FN. Here we use this result as the starting point of a study 
that is based on a novel methodology and on rather extensive statistics. 
A nucleation pattern driven exclusively by native structure (and therefore 
by native topology) is compared with patterns driven by the combined effects 
of protein structure and sequence. If the FN is determined by native 
structure alone the nucleation patterns of different sequences, with 
the same native fold, should be similar and in addition, they should be 
similar to the nucleation pattern of a model whose folding dynamics is 
driven strictly by the structural features of the native fold. \par
 
This paper is organized as follows. The next section describes the
models used and computational methodologies adopted. We
then propose a {\it new} strategy to identify folding nuclei and
present and discuss the simulation results obtained based on it for 
three different model proteins. Finally we draw some conclusions and 
compare our results with those obtained using other strategies and 
simulation efforts.  

\section{Models and methods}

\subsection{Lattice model and simulation details}
\begin{figure*}
{\rotatebox{0}{\resizebox{9cm}{5cm}{\includegraphics{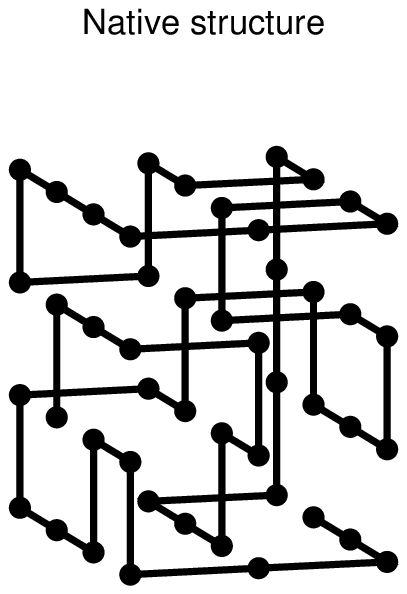}}}} \\
{\rotatebox{0}{\resizebox{5cm}{5cm}{\includegraphics{Fig1b.eps}}}}
\caption{The native conformation used in this study (top) and the 
corresponding contact map (bottom). Each square in the contact map 
represents a non-covalent native
contact, i.e., a contact that is not a covalent linkage. 
\label{fold}}
\end{figure*}

We consider a simple three-dimensional lattice model of a protein molecule 
with chain length $N=48$. In such a minimalist model
amino acid residues, represented by beads of uniform size, occupy the
lattice vertices. The peptide bond that covalently connects amino
acids along the polypeptide chain, is represented  
by sticks with uniform (unit) length corresponding to the lattice spacing 
(Fig \ref{fold}, top).\par
In order to mimic the protein's relaxation towards the native state we use a 
standard  
Monte Carlo (MC) algorithm~\cite{19} together with the kink-jump move 
set~\cite{20}. 
Local random displacements of one or two beads (at the same 
time) are repeatedly accepted or rejected in accordance with the standard 
Metropolis MC rule~\cite{19}.
A MC simulation starts from a randomly generated unfolded conformation 
and the folding dynamics is monitored by following the evolution of the 
fraction of native contacts, $Q=q/L$, where $L=57$ is the number of
contacts in the native fold and $q$ is the number of native contacts
formed at each MC step. The number of MC steps  
required to fold to the native state (i.e., to $Q=1.0$) is the
first passage time (FPT). The native conformation used in this study 
together with its
contact map representation is shown in Figure~\ref{fold}.\par

Unless otherwise specified, folding is studied at the so-called
optimal folding temperature, $T_{opt}$, the temperature that minimizes the
folding time, $t$,~\cite{21,22,23,24,25} which is computed as the mean
first passage  time (MFPT) of 100 simulations.  
This optimal folding temperature may differ from the folding transition 
temperature,
$T_{f}$, at which the probability for finding the protein in an unfolded 
state is the same
as the probability for finding it in the native state. 
In the context of a lattice model $T_{f}$ may be defined as the 
temperature at which the
average value of the fraction of native contacts $<Q>$ is equal to
0.5~\cite{26}. In order to determine $T_{f}$ we averaged $Q$, after
collapse to the native state, over MC simulations lasting at least 20
times longer than the folding time computed at $T_{opt}$.
\par
Protein energetics is modeled using the G\=o and the Shakhnovich models.  

\subsection{The G\={o} model}

In the G\={o} model~\cite{27} the energy of a conformation, defined by the 
set of 
bead coordinates $\lbrace {\vec{r}}_{i} \rbrace$, is given by the contact 
Hamiltonian  
\begin{equation}
H(\lbrace {\vec{r}}_{i} \rbrace)=\sum_{i>j}^N
\epsilon_{ij} \Delta(\vec{r_{i}}-\vec{r_{j}}),
\label{eq:no1}
\end{equation}
where the contact function $\Delta({\vec{r}}_{i}-{\vec{r}}_{j})$, is unity 
if any beads $i$ and $j$ are in contact but not covalently linked, and is zero 
otherwise. 
The G\={o} potential is based on the idea that the native fold is very 
well optimized energetically. Accordingly, it ascribes equal stabilizing 
energies, $\epsilon_{ij}=-1.0$,  to all pairs of beads $i$ and $j$
that form a contact in the native structure, and neutral 
energies, $\epsilon_{ij}=0$, to all non-native contacts. \par 

\subsection{The Shakhnovich model}

By contrast with the G\={o} model, which ignores the protein's chemical 
composition, the Shakhnovich model (see e.g.,~\cite{28}) addresses the 
dependence of protein 
folding dynamics on the amino acid sequence by considering interactions 
between the $20$ different amino acids used by Nature in the synthesis of real 
proteins. Accordingly, the contact Hamiltonian that defines the energy of 
each conformation is given by
\begin{equation}
H(\lbrace \sigma_{i} \rbrace,\lbrace {\vec{r}}_{i} \rbrace)=\sum_{i>j}^N
\epsilon(\sigma_{i},\sigma_{j})\Delta({\vec{r}}_{i}-{\vec{r}}_{j}),
\label{eq:no2}
\end{equation}
\noindent
where $\lbrace \sigma_{i} \rbrace$ represents an amino acid sequence,
and $\sigma_{i}$ stands for the chemical identity of bead $i$. 
In this case both the native and the non-native contacts contribute 
energetically to the folding process. The interaction 
parameters $\epsilon$ are taken from the $20 \times 20$
Miyazawa-Jernigan matrix, derived from the distribution of contacts of
native proteins~\cite{29}.\par
Two non-homologous sequences, numbered 1 and 2, were studied within the 
context of
the Shakhnovich model. The latter were designed to fold into the native
conformation shown in Figure \ref{fold}  with the method developed by
Shakhnovich and Gutin based on random heteropolymer theory and
simulated annealing techniques~\cite{30}.\par
Table \ref{tabdata} summarizes some kinetic and thermodynamic properties 
of the model proteins discussed above.

\begin{table*}
\begin{tabular}{|c | c c c c|}
\hline
Sequence      & E$_{nat}$&       T$_{opt}$  & T$_{f}$ & $\log_{10}({t})$  \\ 
\hline \hline
 G\={o}  &  $-57.00$  &  0.65  &0.770&    5.95 $\pm$ 0.03 \\
  $1:$
\texttt{EPEWQLEFDNSNYAWPANYAQHLPGMYRFTVFDMQRNHTSCKLCFLFS}
  &    $-24.34$  & 0.29 & 0.305  &   6.84 $\pm$ 0.04 \\
  $2:$  
\texttt{CIFDLEFECPAFPAPIGWLGLVSVVYLFPVRYCRLCMFNCRFKTKTRC}
& $-26.84$  &      0.32  & 0.332  &   6.53 $\pm$ 0.04\\  
\hline
\end{tabular}
\caption{Kinetic and thermodynamic properties of the three model proteins. 
The
  folding time, $t$, is measured at the optimal folding
  temperature, $T_{opt}$. Also shown is the folding transition
  temperature, $T_{f}$, and the native state's energy
  $E_{nat}$. \label{tabdata}}  
\end{table*}

\section{A general strategy to identify the folding nucleus}

We define the FN as a {\it specific}  set of native 
contacts which, once formed, prompts rapid and highly probable folding to 
the native state. 
In what follows we render a methodology to investigate 
the existence of folding nuclei in the folding of 48-mer lattice polymers 
whose energetics are modeled by the G\={o} or by the MJ potential.  
\par
The vast majority of small (i.e. with less than 100 amino acids), single 
domain proteins fold in a two-state manner with a relaxation rate following 
single-exponential kinetics~\cite{31}. 
Two-state folding is often rationalized through a `classical' mass-action 
scheme~\cite{32}. Accordingly, the ensemble of conformations that make up 
the unfolded state ($U$) is separated from the native fold ($N$) by a free 
energy barrier along some appropriately defined reaction coordinate. The 
ensemble of conformations that lie on the top of the reaction barrier is 
the so-called transition state (TS). By definition, TS's conformations have 
folding probability $P_{fold}=1/2$ (in other words, TS's conformations have 
a probability 0.5 to fold before they unfold)~\cite{33}. 
If folding occurs via nucleation, conformations that rapidly reach 
the native state with high probability $P_{fold}\gg 1/2$ are post-transition state conformations in which the FN is formed. The latter is indeed a postcritical FN since its formation inevitably leads to the formation native state~\cite{4}. In the present study we are therefore interested in postcritical folding nuclei. An appropriate structural analysis of a significantly large ensemble of such conformations should therefore reveal, with a high degree of statistical confidence, a set of common contacts which is the FN. 
To build such an ensemble we consider 1000 different folding events and, 
for each individual event, we identify the earliest formed conformation (EFC) 
that folds rapidly and with high probability, $P_{fold}\geq P^*_{fold}$. 
In order to determine the EFC for a given folding event conformations are 
sampled at times 
\begin{equation}
t_s(n)={\rm FPT}-n\Delta t ,
\label{time}
\end{equation}
where $\Delta t$ is an appropriate sampling interval and $n=1, 2, ...$. 
More precisely, starting with $n=1$, the folding probability, $P_{fold}$, of 
the conformation collected at time $t_{s}(1)$ is computed; this amounts to
determining the fraction of folding simulations (in a set of 100 MC runs) 
which, starting from that conformation reach the native state without 
passing through conformations with $Q<Q_{U}$, i.e., the protein folds 
before it unfolds (we consider a protein to be unfolded if its fraction of 
native contacts is smaller than some cut-off $Q_{U}$). If $P_{fold}<P^*_{fold}$ the 
conformation is discarded. Otherwise, if the folding time $t$ is smaller than some cut-off time 
$t_{max}$, the procedure described above is repeated for $n=2$ etc. The 
EFC for a given folding event is the conformation corresponding to the 
largest $n$ which has $P_{fold}\geq P^*_{fold}$ and $t<t_{max}$. In the following section we discuss in some detail the procedure used to fix the parameters $Q_{U}$, $\Delta t$ and $t_{max}$.

\subsection{Nucleation in the G\=o model}
\subsubsection{Determination of $Q_{U}$,  $t_{max}$ and $\Delta t$}

While it is trivial to identify the native state (since is the unique
conformation with $Q=1.0$) it is not straightforward to decide weather a
conformation belongs to the ensemble of unfolded conformations or is
kinetically close (i.e., rapidly converts) to the native state.\par 
The fraction of native contacts $Q$ has been extensively used in simulation 
studies as a reaction coordinate, i.e., as a parameter that quantifies the
degree of folding \cite{26, pande, sali, socci}. 
In general, however, $Q$ measures closeness to the native structure in
energetic (or thermodynamic) terms only. It has been argued that,
unless the energy landscape is considerably smooth, thermodynamic
closeness does not necessarily imply kinetic proximity to the native
structure~\cite{34}. However, 
even if
the suitability of $Q$ as a reaction coordinate is questionable, very
small $Q$s must necessarily identify unfolded conformations (i.e.,
that are thermodynamically and kinetically distant from the native fold ).\par
\begin{figure}
\begin{center}
{\rotatebox{0}{\resizebox{8cm}{5cm}{\includegraphics{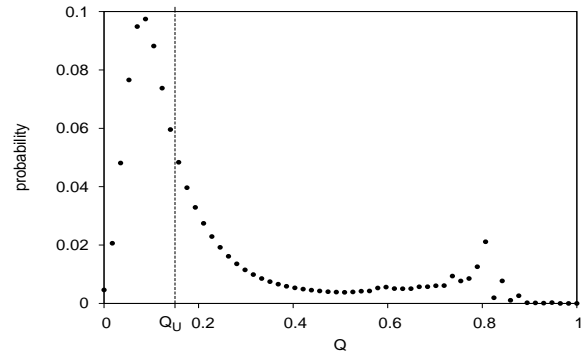}}}}
\end{center}
\caption{Probability of finding a conformation with
  fraction of native contacts $Q$ a function of $Q$. $Q_U$ is the
  fraction of native contacts below which the protein is considered to be 
  unfolded. The probability of $Q=1.0$ vanishes since the 
  simulation stops when the protein reaches the native state. 
\label{free}}
\end{figure}

  
In order to distinguish unfolded conformations from other conformers
we have computed the probability of finding a conformation with
a fraction of native contacts $Q$ as a function of $Q$ in a sample of
200 different folding events. Two peaks are apparent in the
graph reported in Figure~\ref{free}: a high-probability peak centered
at $Q=0.088$ and another one, of considerably lower probability, that 
appears immediately prior to the native fold. The high probability peak is 
clearly associated with the unfolded states. The cut-off $Q_{U}$ is chosen 
such that more than half of the unfolded peak lies to the left of $Q=Q_{U}$. 
In what follows we take $Q_U=0.15$ but note that other values of 
$Q_{U}$ were tested and were found to lead to the same results. \par 
The probability for the protein to be in high-$Q$ conformations is 
small but non negligible (Fig \ref{free}). This happens because 
the optimal folding temperature $T_{opt}$, at which data was collected, is 
well below the system's folding transition temperature $T_{f}$
(Table \ref{tabdata}). Accordingly, the protein may be trapped in low 
energy
conformations that share a high degree of structural similarity with
the native fold (i.e., whose fraction of native contacts is $Q\sim0.8$).

By definition, the formation of the FN prompts rapid and highly probable
folding ($P_{fold}\geq P^*_{fold}$). The cut-off parameter $t_{max}$ 
(i.e., the maximum number of MC steps in which the protein is required to 
reach the native fold) is therefore a particularly important step of the procedure
proposed to identify the FN.\par

\begin{figure}
\begin{center}
{\rotatebox{0}{\resizebox{8cm}{5cm}{\includegraphics{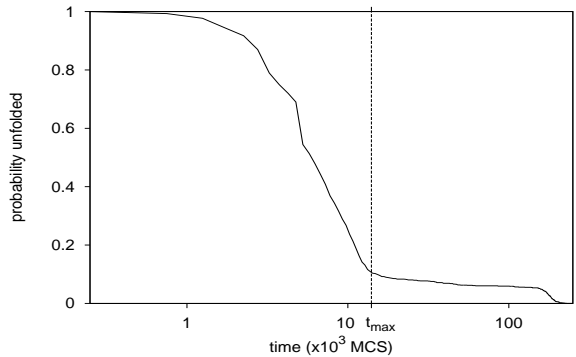}}}}
\end{center}
\caption{Fraction of unfolded conformations as a
 function of time (in a $\log_{10}$-scale). About 90\% of the
 conformations fold in $1.4$x10$^4$ MCS while the remaining 10\% 
 fold in times that are about one order of magnitude longer.  
\label{ftime}}
\end{figure}

A tentative sampling interval (about 2 orders of magnitude smaller than 
the folding time for this model protein) was used to collect an ensemble 
of $\sim 2000$ conformations with $P^*_{fold}=1$ from 100 different folding 
events. The vast majority ($>90$\%) of such conformations were found to 
reach the native state in time $t$ less than $1.4$x$10^4$ MCS while about 
10\% take a considerably longer time to fold (Figure \ref{ftime}).  \par 

Two (folding) time scales are clearly distinguished in this ensemble
of conformations. The shorter time scale corresponds to conformations where 
the FN has the highest probability of being formed, while the longer one 
is associated with folding events during which the protein is trapped in 
low energy states which, despite despite sharing a large similarity
with the native fold, do not have the FN formed (Figure~\ref{free}).
In order to eliminate the latter conformations $t_{max}$ was set to 
$1.4$x$10^4$ MCS. \par 

The efficiency of the sampling procedure may be improved by choosing the 
sampling interval, $\Delta t$, appropriately. Let $\rm FPT-FPT_{\rm EFC}$ 
be the number of MC steps required to complete folding once the EFC forms 
at time $\rm FPT_{\rm EFC}$ in a given folding event. We define
$t_{\rm EFC}$ as the average folding time of the EFC of 100 
folding events (i.e., $t_{\rm EFC}$ is the average of $\rm FPT-FPT_{\rm 
EFC}$ computed over 100 folding events). Ideally, the 
sampling interval should be smaller than $t_{\rm EFC}$, or at least of the 
same order of magnitude. In practice, for a tentative  
$\Delta t$, we compute $t_{\rm EFC}$ by averaging $N\Delta t$ in 100 
folding events where $N$ is the maximum value of $n$ for each event.  
We fix $\Delta t$ if the corresponding $t_{\rm EFC}$ lies between 
5$\Delta t$ or 10$\Delta t$.
For the model protein considered in this section we have found that 
$t_{\rm EFC}\sim 6000$ for $\Delta t=1000$ MCS, which means that, on average, 
the EFCs are collected at a sampling time $t_s(6)$.\par
\begin{figure}
\begin{center}
{\rotatebox{0}{\resizebox{8cm}{5cm}{\includegraphics{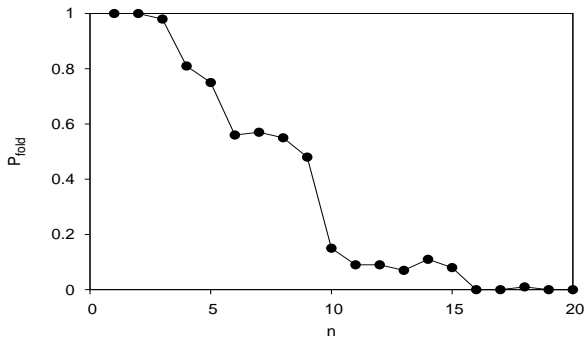}}}}
\end{center}
\caption{A typical plot of $P_{fold}$ as a function of $n$ 
(with $\Delta t=1000$).
Conformations collected at small $n$ have a very high $P_{fold}$ 
and some of them have $P_{fold}=1$. 
\label{prob}}
\end{figure}

   
In Figure \ref{prob} the dependence of $P_{fold}$ on $n$ is shown
for a single folding event. The folding probability is zero when
$n=16$ but as time approaches the FPT (i.e. for $n<16$) the protein
explores a series of conformations, with $P_{fold}\neq 0$ and reaches
the native state with $P_{fold}=1$ when $n=0$. The conformations 
corresponding to $n=1$ and $n=2$ have $P_{fold}=1$ as well and reach the 
native state in time $t_f<t_{max}$. Thus, the EFC for this folding event 
is the conformation which corresponds to $n=2$.

\subsubsection{A folding nucleus determined solely by native topology}

Having fixed the parameters $Q_{U}$, $\Delta t$ and $t_{max}$ we ran 1000 
different folding events from which an ensemble of 1000 conformations (one
conformation per folding run) were collected. The latter are all EFCs,
i.e., the earliest conformations in folding events that collapse rapidly 
to the native state (i.e., their folding time is $t<t_{max}=14000$ MCS) with 
unit folding probability. The average fraction of native contacts of this 
ensemble of conformations is $<Q>_{\rm EFC}=0.67$. \par   
We start by labeling the 57 native contacts as in Table \ref{t1}.

\begin{table*}
{\footnotesize
\begin{center}
\begin{tabular}{||c|r@{ : }l||c|r@{ : }l||c|r@{ : }l||c|r@{ :
    }l||c|r@{ : }l||}
\hline
{\bf Contact} & R$_i$ & R$_j$ & {\bf Contact} & R$_i$ & R$_j$ & {\bf Contact} & R$_i$ & R$_j$
& {\bf Contact} & R$_i$ & R$_j$ & {\bf Contact} & R$_i$ & R$_j$ \\ \hline\hline
0 &0 &41 &12 &6 & 35&24 &21 &26 &35 &32 &35 &46 &5 &44 \\\hline
1 &7 &44 &13 &23 &26 &25 &5 &42 &36 &1 & 20&47 &14 &33 \\\hline
2 &10 &47 &14 &27 &34 &26 &6 &41 &37 &2 &21 &48 &15 &34 \\\hline
3 &11 &32 &15 &28 &33 &27 &7 &40 &38 &3 &22 &49 &17 &36 \\\hline
4 &12 &33 &16 &0 &35 &28 &8 &39 &39 &4 & 23&50 &18 &37 \\\hline
5 &14 &25 &17 &1 &34 &29 &9 &38 &40 &6 &27 &51 &24 &29 \\\hline
6 &15 &26 &18 &2 &27 &30 &11 &36 &41 &8 &35 &52 &25 &28 \\\hline
7 &17 &34 &19 &4 &29 &31 &12 &17 &42 &9 &36 &53 &28 &31 \\\hline
8 &40 &43 &20 &5 &30 &32 &13 &16 &43 &0 &39 &54 &30 & 45\\\hline
9 &0 &37 &21 &6 &31 &33 &15 &20 &44 &2 &41 &55 &31 & 46\\\hline
10 &1 &18 &22 &7 &46 &34 &16 &19 &45 &3 &42 &56 &32 &47 \\\hline
11 &4 &27 &23 &8 &47 \\\cline{1-6}
\end{tabular}
\end{center}
}
\caption{For structures that like ours are maximally compact cuboids
with $N=48$ residues there are 57 native contacts. This table displays 
the correspondence between the contact number 
and the pair of residues involved in each contact. 
\label{t1}}
\end{table*}

\begin{figure}
\begin{center}
{\rotatebox{0}{\resizebox{8cm}{5cm}{\includegraphics{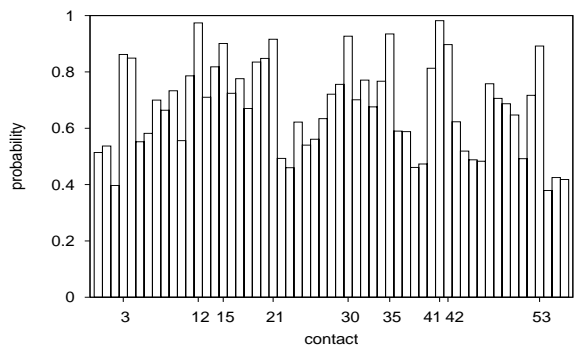}}}}
\end{center}
\caption{The contact histogram for the G\={o} model showing, for each native 
contact, the probability of being formed in the ensemble of 1000 EFC 
conformations that fold in time $t<1300$ MCS with unit folding probability. 
The nine contacts identified by number have the highest probability (i.e. 
probability $>85$\%) of being formed.
\label{hist}}
\end{figure}

\begin{figure*}
{\rotatebox{0}{\resizebox{5cm}{5cm}{\includegraphics{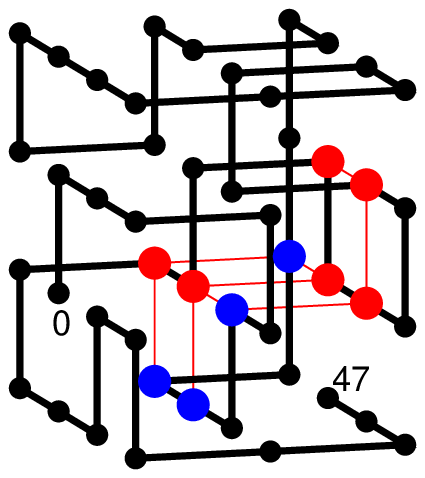}}}}
{\rotatebox{0}{\resizebox{5cm}{5cm}{\includegraphics{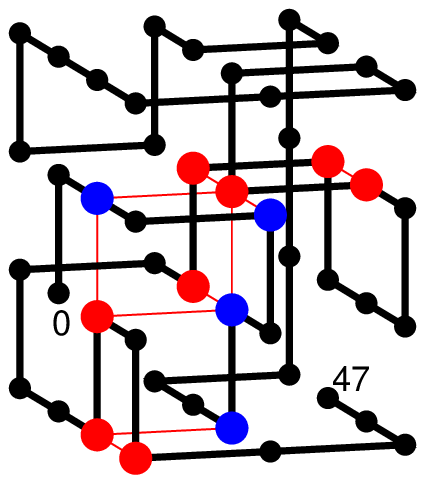}}}}
{\rotatebox{0}{\resizebox{5cm}{5cm}{\includegraphics{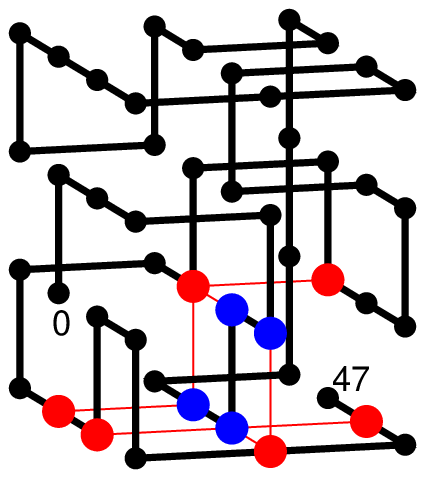}}}}
\caption{The folding nucleus for the G\=o model (left), for sequence 1 
(center) and for sequence 2 (right) is the set of 9, 10 and 8 contacts, 
respectively, colored in red. Residues whose number along the sequence is less than 12 are 
colored in blue and those whose number along the sequence is larger than 26 are 
colored in red.}     
 \label{nat} 
\end{figure*}

For each native contact we define the contact probability as the number 
of conformations in
which the contact is formed normalized to the total number of
conformations in the sample. Results reported in Figure \ref{hist}
show that the contact probability varies considerably among the 57
native contacts, an observation that is particularly evident
for probabilities larger than 50\%. This finding strongly suggests
that, while the establishment of some contacts (e.g., 12 and 41, which
are present in over 95\% of the conformations analyzed) is an
essential requirement to ensure rapid folding, the formation of others 
(e.g., 2 and 54 which appear with probability $<$40\%) does not 
appear to be a requisite to fast folding. The set of 9 contacts, 
involving 
residues 6, 8, 9, 11, 28, 31-33, 35, and 36 (Figure \ref{nat}, left), and 
identified by contact number in Figure \ref{hist}, seems to be particularly 
relevant. Indeed, each individual contact is formed in more than 85\% of 
the conformations analyzed, and all of the 9 contacts are {\it 
simultaneously} formed in 64\% of the conformers. Moreover, on average, 
$8.2$ of them are present in the ensemble of conformations considered.\par 
The fact that rapid folding is associated with the formation of a set of 
highly probable contacts suggests that such a contact set is the 
FN. \par

There is, of course, a
certain degree of arbitrariness in the choice of the probability cut-off
that is used to identify the highly probable contacts, and therefore the set of
contacts identified above is a {\it putative} FN.

\subsection{Nucleation in the Shakhnovich model}

In order to investigate the importance of amino acid sequence in the
formation of the FN, we studied the folding of two
non-homologous sequences (numbered 1 and 2) (Table \ref{tabdata}).

\subsubsection{Determination of $Q_{U}$, $\Delta t$ and  $t_{max}$}

\begin{figure}
{\rotatebox{0}{\resizebox{8cm}{5cm}{\includegraphics{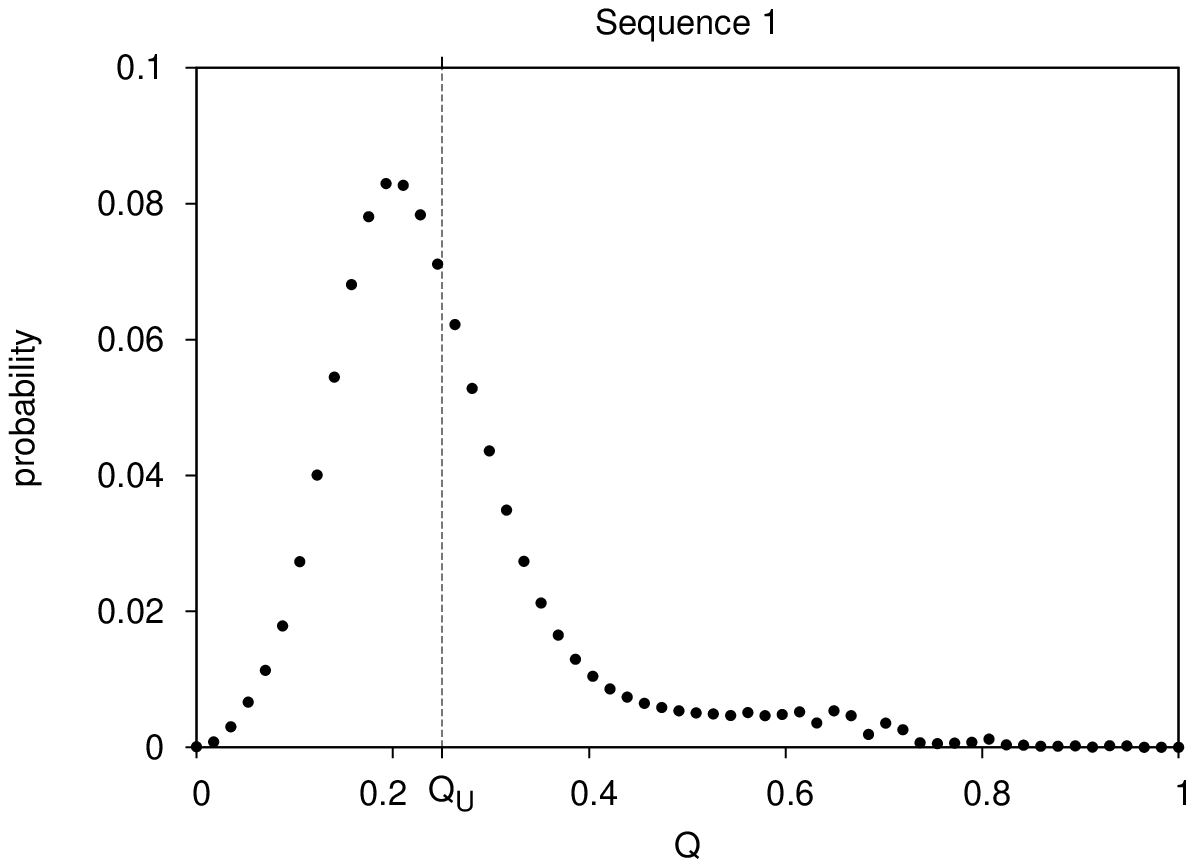}}}} \\
{\rotatebox{0}{\resizebox{8cm}{5cm}{\includegraphics{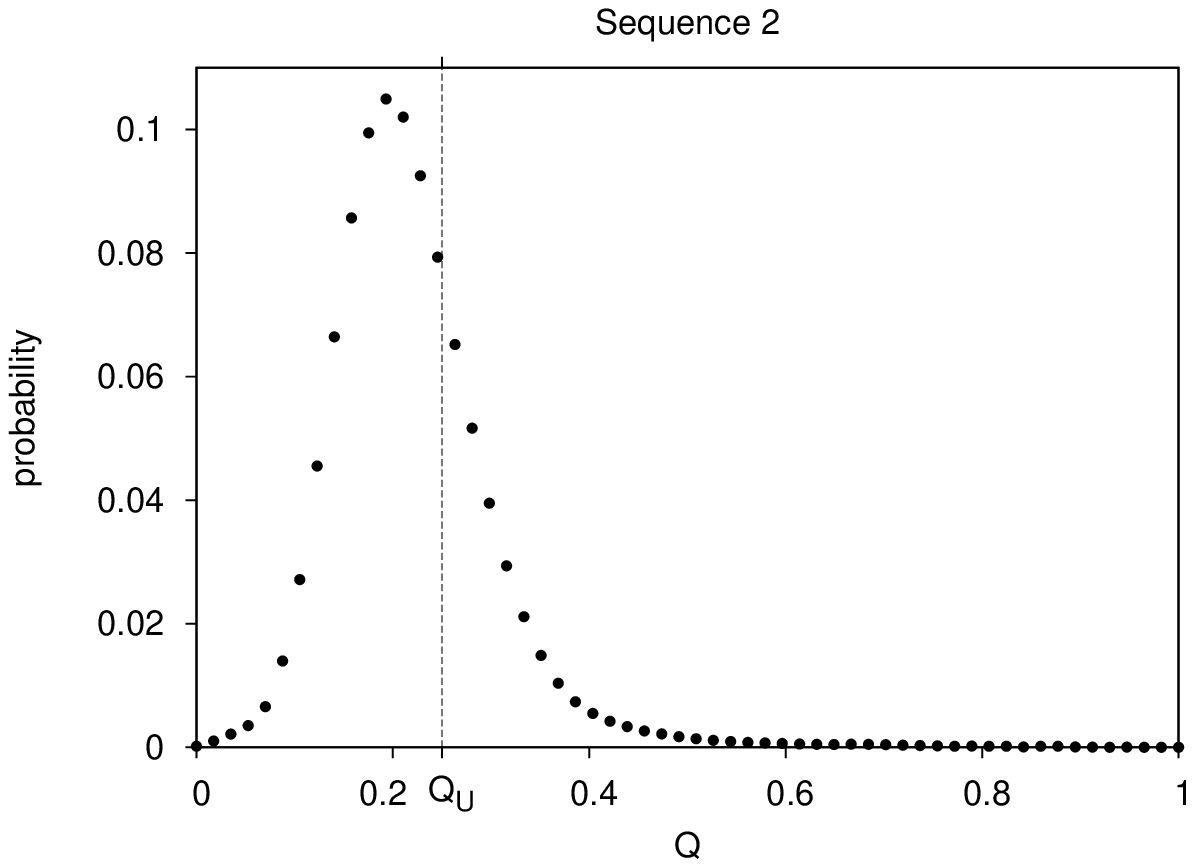}}}}
\caption{Probability of having a conformation with fraction of native contacts
  $Q$  for sequences 1 (top) and 2 (bottom). The peak at small $Q$ 
  is well defined for both model proteins. The probability curve for sequence 2 
falls sharply to zero as $Q$ increases, while for sequence 1 there is a small 
probability 
for the system to be found in conformations with $0.5<Q<0.7$. In either case the 
protein is 
considered to be unfolded when the fraction of native contacts is smaller than 
$Q_U=0.25$.  
\label{MJprob}}
\end{figure}

In the G\={o} model the so-called topological frustration~\cite{36}
results from polymer properties of the chain such as
connectivity~\cite{9,33}, excluded volume effects, and quirks of the
native topology, such as lack of symmetry~\cite{37}. Topological
frustration is the only type of frustration in models which, like the 
G\={o} model, are native centric. On the other hand, by taking into account the
protein chemistry, the Shakhnovich model also exhibits energetic frustration. 
The latter typically leads to longer folding times and, at temperatures below the folding
transition temperature, the chain is prone to get trapped in low energy states~\cite{37}. 
This implies that, in contrast with the G\={o} model, for which  $T_{opt}$ is well below 
$T_{f}$, the two Shakhnovich protein sequences have optimal folding 
temperatures which are close to the system's folding transition temperatures (Table
\ref{tabdata}). Thus, although the observed folding times are longer than
those found for the G\=o model (Table \ref{tabdata}), the Shakhnovich model proteins do 
not get trapped in high-$Q$, low energy states. Indeed, the $Q$ 
probability distributions are not peaked in the high-$Q$ ($Q \sim 0.8$)
region (Figure \ref{MJprob}) although both models exhibit a well defined, high-probability 
low-$Q$ peak, at $Q=0.20$, corresponding to the unfolded states. Applying the same 
criterion for the choice of cut-off $Q_{U}$, one considers a conformation unfolded if 
$Q<Q_U=0.25$. As before, we have found that the results for the FN are robust with respect 
to small variations in the choice of $Q_{U}$.

\begin{figure}
{\rotatebox{0}{\resizebox{8cm}{5cm}{\includegraphics{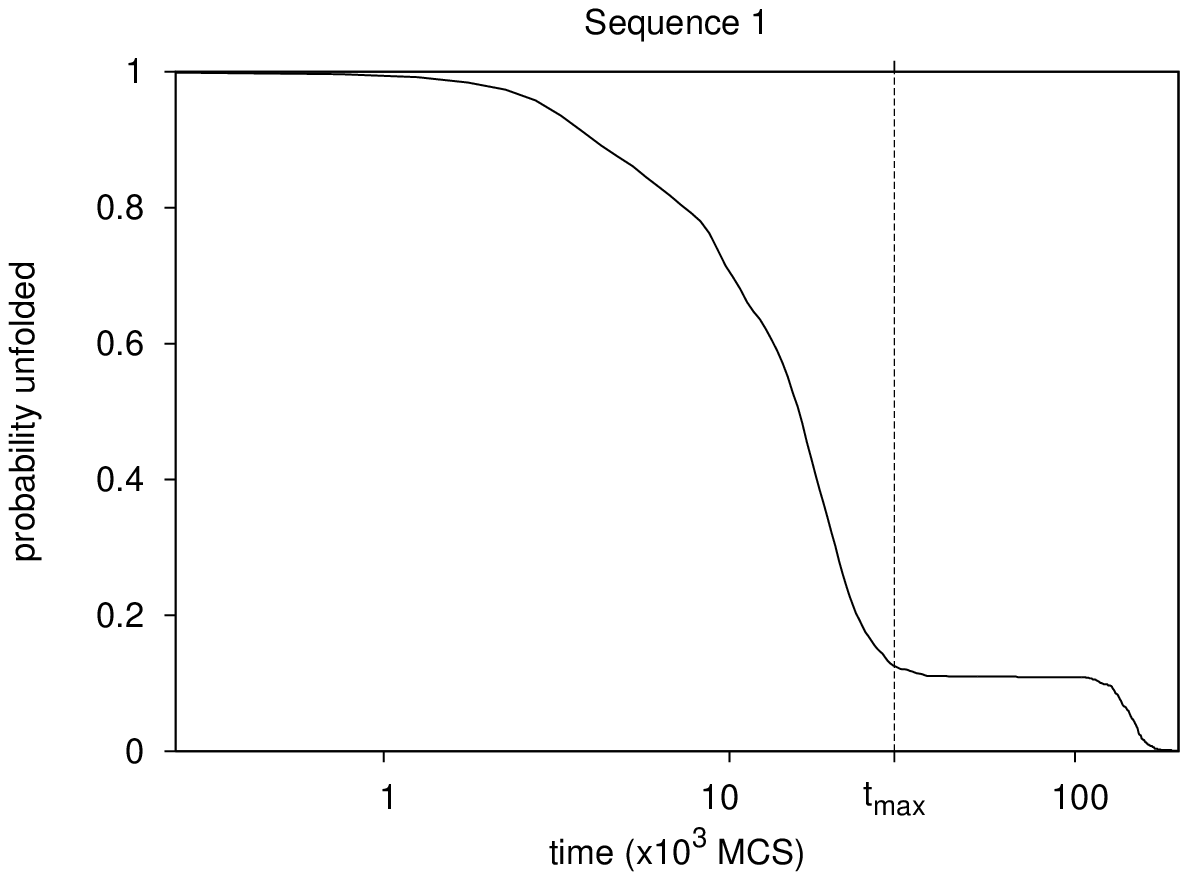}}}}\\
{\rotatebox{0}{\resizebox{8cm}{5cm}{\includegraphics{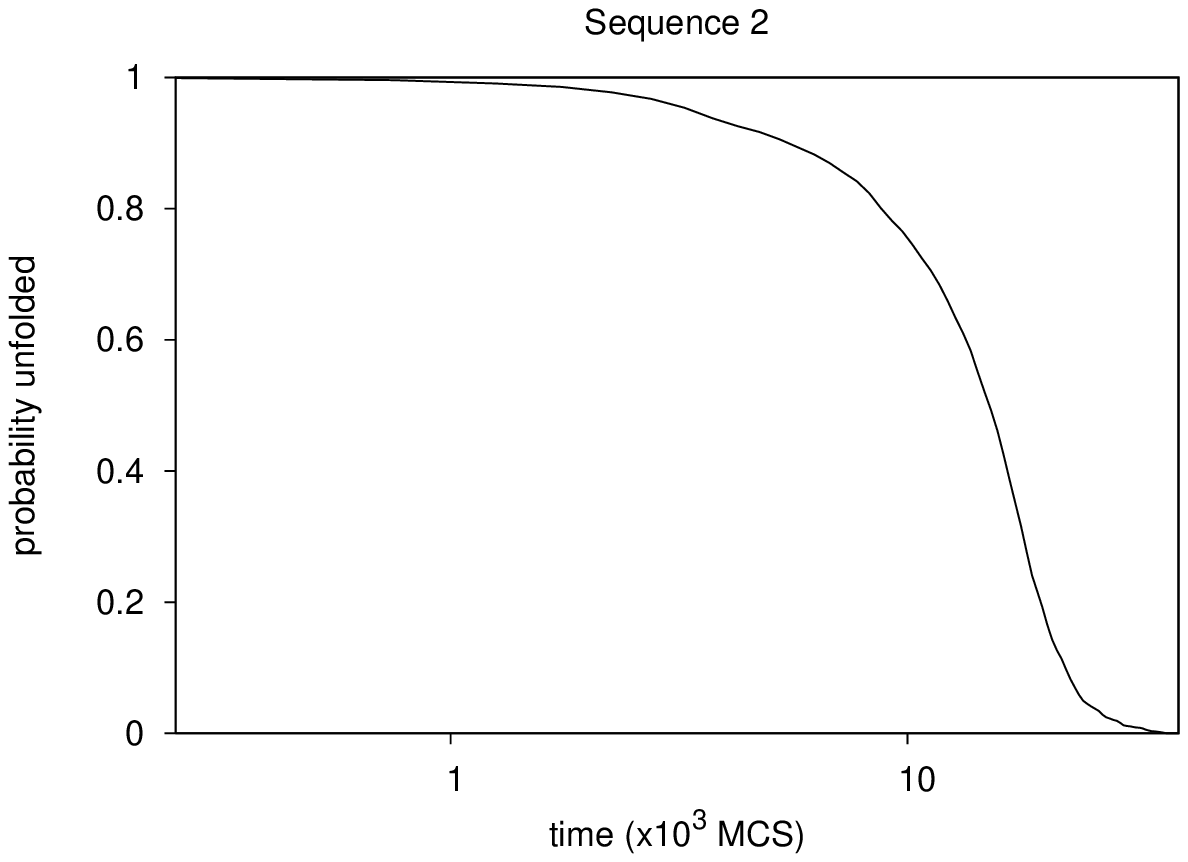}}}}
\caption{Fraction of unfolded conformations as a function of
  time starting from conformations with $P_{fold}>0.90$. For sequence 1 (top) 
two time-scales, differing by one order of magnitude, may be observed. In order 
to select the fastest folders the cut-off time is fixed at $t_{max}=30000$ MCS. 
For sequence 2 (bottom) there is no need to use a cut-off time since all folding  
are of the same order of magnitude. 
\label{MJtime}}
\end{figure}

In order to fix $t_{max}$, a set of $\sim$ 1200 conformations (per sequence),
with $P^*_{fold}=0.90$, is collected from 100 different folding events
and the corresponding folding times are measured. For sequence 1, two folding 
time scales are observed (Figure \ref{MJtime}, top). 
The fraction of native contacts in the ensemble of sequence 1's
conformations is $Q=0.72\pm 0.12$. Since there is a small
probability for sequence 1 to be in conformations with $Q\sim$ 0.7
(Figure \ref{MJprob}) the longer time-scale may be ascribed to the population
of these relatively high-$Q$ conformations which, being local
energy minima, will slow down folding. In order to disregard these
conformations the cut-off time is set to $t_{max}=30000$ MCS. 
By contrast, for sequence 2 the folding times are all of the same order 
of magnitude (Figure \ref{MJtime}, bottom) and there is no need to use a 
cut-off time, $t_{max}$.\par 
The reason for taking $P^*_{fold}=0.9$, instead of
$P^*_{fold}=1.0$ as in the Go model, is that
the latter leads, in the Shakhnovich model, to an ensemble of
conformations with a high average fraction of native contacts ($<Q> \sim 0.85$). 
The latter are practically folded and thus are not suitable to distinguish 
the contacts that belong to the FN from other trivial contacts. 

To improve the efficiency of the sampling procedure we have, also for the Shakhnovich 
model proteins, optimized the sampling intervals as described previously. We have found 
that $\Delta t=1000$ MCS works well for both proteins yielding $t_{\rm EFC}\sim9500$ MCS 
and $t_{\rm EFC}\sim14000$ MCS for sequence 1 and 2 respectively, i.e., on average the 
EFCs for sequence 1 are collected at $t_s(10)$ while for sequence 2 they are collected 
at $t_s(14)$.  

\subsubsection{Folding nuclei determined by topology and
  protein sequence} 

Two ensembles, each comprising 1000 EFCs, were obtained for sequences
1 and 2 using the parameters discussed in the previous section, with 
$<Q>_{EFC}=0.65$ and $<Q>_{EFC}=0.62$ for sequences 1 and 2 respectively. 
These values of $<Q>$ are similar to that of the G\=o 
model and considerably lower than those obtained if $P^*_{fold}=1.0$ is used
for the Shakhnovich model, allowing the distinction of the contacts in a 
putative FN from other spurious contacts.\par
The native structure of sequences 1 and 2 is the same as that of the 
G\={o} model and the same numbering of native contacts is used (Table \ref{t1}).\par

\begin{figure}
{\rotatebox{0}{\resizebox{8cm}{5cm}{\includegraphics{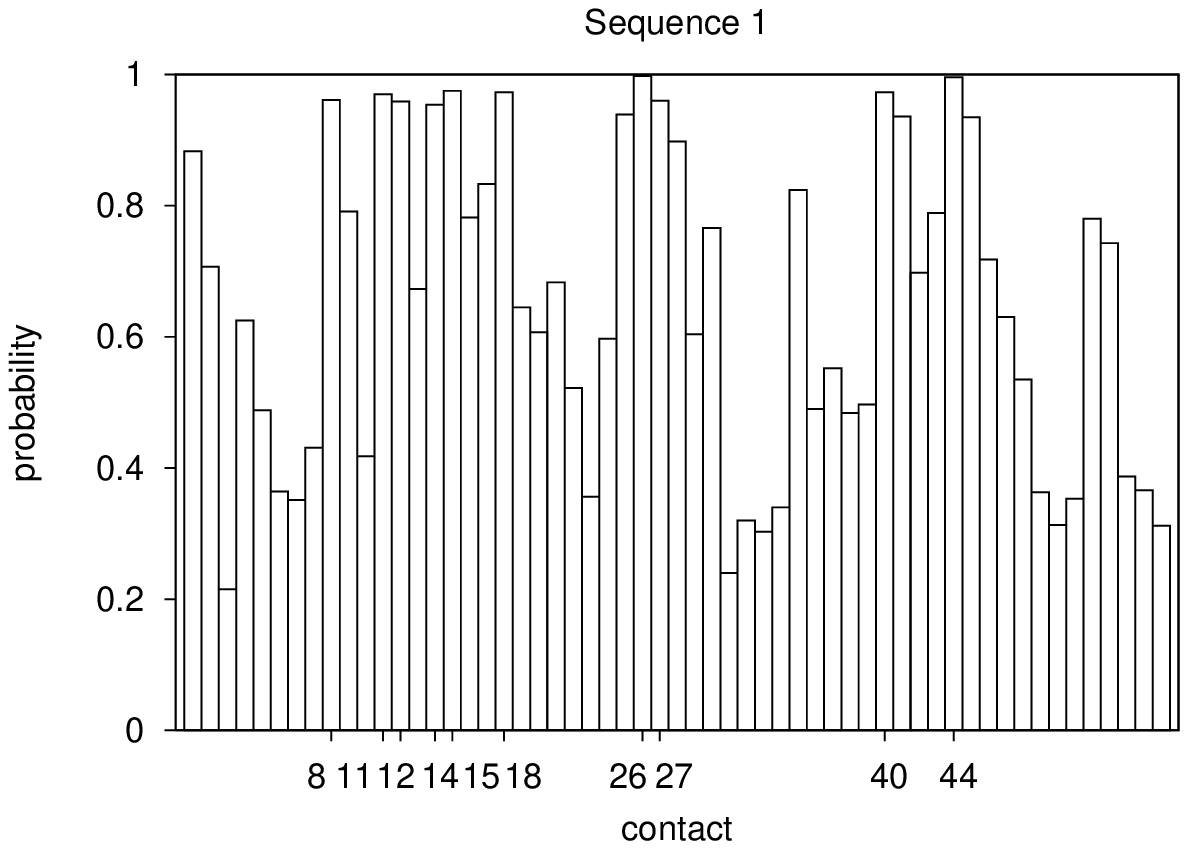}}}} \\
{\rotatebox{0}{\resizebox{8cm}{5cm}{\includegraphics{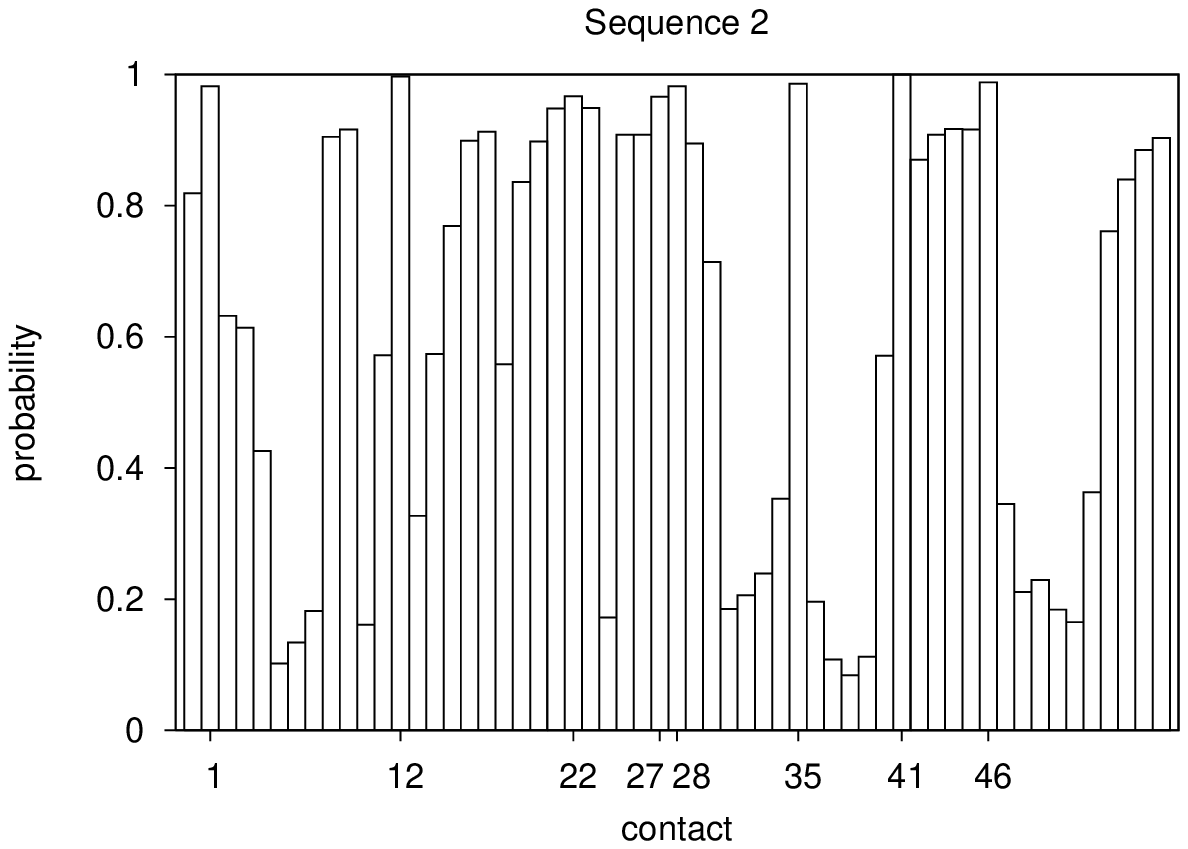}}}}
\caption{Contact histograms for sequence 1 (top) and sequence 2 (bottom).
Contacts present with the highest probability ($>95\%$) are identified
by contact number.  
\label{histmj}}
\end{figure} 

From the analysis of the contact histograms we observe that some native contacts 
are present with very high probability ($>95\%$)(Figure
\ref{histmj}). We consider the {\it putative} FN as the set of the most
probable contacts.\par
For sequence 1 the FN is thus formed by 10 native linkages (identified by contact 
number in Figure~\ref{histmj}, top) involving 12 residues 
(namely, 2, 4, 6, 7, 27, 28, 33, 34, 35, 40, 41, and 43)(Figure~\ref{nat}, center). 
The 10 contacts forming the FN are {\em simultaneously} present in 82\% of the EFC 
conformations analyzed and, on average, the latter have 9.7 of these contacts formed. 
It is interesting to note that the average stability of the contacts forming the FN 
is 62\% higher than the average stability of the 57 native contacts of the folded 
protein (Table \ref{t2}).
For sequence 2 the FN is formed by 8 native contacts (identified by contact number 
in Figure~\ref{histmj}, bottom) and 10 residues (namely, residues 5, 6, 7, 8, 32, 35, 
39, 40, 44, and 46) (Figure \ref{nat}, right). 
The 8 contacts forming the FN are {\em simultaneously} present in 90\% of the EFC 
conformations analyzed and, on average, the latter have 7.9 of these contacts formed. 
In this case the average stability of the FN's contacts is 53\% higher than the
average stability of the protein's native contacts (Table \ref{t2}). 
\par 
The two folding nuclei have 2 native contacts (12 and 27) and 4 residues in common. 
These native contacts are non-local linkages between residues 6 and 35 and between 
residues 7 and 40, suggesting that the establishment of the corresponding long range 
interactions might be determinant to ensure rapid folding.\par 

Structurally speaking the FN of sequence 1 consists of two loops,
one
formed by residues 2, 27, 41, and 6 and the other by residues, 41, 6, 40,
and
7 (Figure \ref{nat}, center). Each of these loops is formed by contacts
located in the interior of the protein, while in sequence 2 a significant
fraction of the FN's contacts are located on the fold's surface (Figure
\ref{nat}, right).

\subsection{Nucleation scenarios and contact stability}

The G\={o} FN shares 22\% of its contacts with sequence 1 and 33\% with sequence 2. 
The presence of these contacts in the folding nuclei of the Shakhnovich models is 
driven by native topology. Indeed, the average stability of the Shakhnovich contacts 
that are also present in the G\={o} model is up to 25\% lower than the average 
stability of the remaining contacts in the FN (Table~\ref{t2}, columns 3 and 4) but 
they are formed with equally high probability $>95$\%.\par
The extremely high probability ($\sim$1) of the contact between residues 6 and 35 
(i.e. contact 12 in the contact histograms) in all the three model proteins is a 
robust feature of the nucleation mechanism. Another interesting observation regarding 
these residues is that they make-up a network of 7 native contacts in the fold (whose 
average range is 25 units of backbone distance) and about half of these contacts are 
present in each FN which suggests that they might be key residues in the folding process. 
We have performed exhaustive single-point mutations in all of the 48 residues 
and, in agreement with the above hypothesis, we have found that two mutations, one on residue 6, 
and the other on residue 35, lead to the largest increases in folding times (the folding time 
increases by up to 6-fold with respect to that of the wild-type sequence)~\cite{38}. \par

The average stability of the G\={o} FN's contacts that do not participate 
in the Shakhnovich folding nuclei, of sequences 1 and 2, is up to 66\% lower than the 
protein's 57 native contacts (Table~\ref{t2}, columns 1 and 5). By contrast, the 
contacts that are exclusive to the Shakhnovich folding nuclei are up to 90\% more stable 
than the protein's 57 native contacts (Table~\ref{t2}, columns 1 and 4). Moreover, as we 
have already pointed out, the Shakhnovich folding nuclei are up to 81\% more stable than 
the protein's 57 native contacts (Table~\ref{t2}, columns 1 and 2).\par

\begin{figure}
\begin{center}
{\rotatebox{0}{\resizebox{8cm}{5cm}{\includegraphics{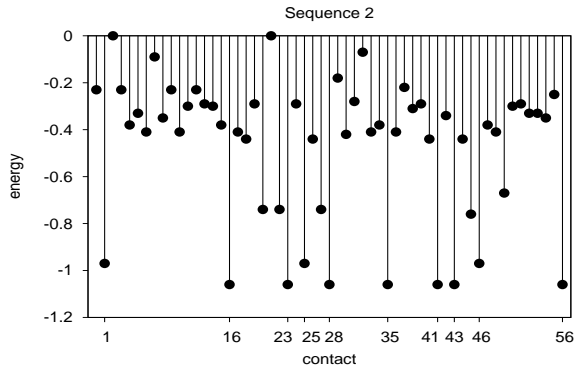}}}}
\end{center}
\caption{Energy of each native contact. Half of the most stable contacts, identified in the figure by contact number, are present in sequence's 2 folding nucleus. 
\label{inthist}}
\end{figure}

Clearly, by ascribing different stabilities to the protein's native contacts, the protein 
sequence promotes an overall change of the nucleation scenario, which in the G\={o} model 
is driven solely by the topological features of the native fold. To see how this happens 
in more detail we investigated the effect of contact stability in the contact histogram 
(i.e. in the determination of the FN) of sequence 2. 
The most stable contacts in this case are contacts 1, 16, 23, 25, 28, 35, 41, 43, 46, 56 
(Figure~\ref{inthist}) and, not surprisingly, half of them belong to the FN (Figure~\ref{histmj}). 
It is interesting to note that, by being particularly stable, some contacts may indirectly 
promote an increase in the probability of occurrence of other less stable contacts. This 
feature is well illustrated by residue 47 and the three contacts it establishes in the fold. 
The latter appear with considerably high probabilities in the contact histogram. 
The probabilities of contacts 23 and 56 (which are considerably lower in the G\={o} model) may 
be ascribed to their very high stabilities. However, contact 2 is a neutral one and, in spite 
of its relative low stability, its probability is higher when compared with other stable 
contacts in the protein. This presumably happens because the very high stability of contacts 
23 and 56 forces residue 47 to be in its native environment (i.e. to have all of its native 
contacts formed simultaneously) which naturally increases the probability with which contact 
2 is formed.\par    

Stability is indeed a considerably determinant factor for the Shakhnovich FN,
but is not the whole story. The presence of G\=o contacts in the nucleus,
is not energetically favorable (Table \ref{t2}, columns 2 and 3), but is very
relevant from a functional point of view as discussed in the next section.

\begin{table*}
{\footnotesize
\begin{center}
\begin{tabular}{|c|c|c|c|c|c|}
\hline
\multicolumn{6}{|c|}{\rule[-2mm]{0mm}{6mm}\small{\bf Mean 
    energy per contact}} \\
\hline\hline
 & protein & S$_\mathrm{FN}$ & S$_\mathrm{FN}$ $\wedge$ 
 G\=o$_\mathrm{FN}$ &  S$_\mathrm{FN}$ $\wedge$ ($\sim$
 G\=o$_\mathrm{FN}$) &  ($\sim$ S$_\mathrm{FN}$) $\wedge$ G\=o$_\mathrm{FN}$\\
\hline
\hline
Sequence 1  & -0.427  &-0.691 &  -0.579  & -0.719   &   -0.424\\
\hline
Sequence 2 & -0.471  & -0.854 &  -0.783  &  -0.896  &    -0.283\\
\hline
\end{tabular}
\end{center}
}
\caption{Mean energy per contact in different contact sets. In the the first column the average 
is computed over the protein's 57 native contacts. S$_\mathrm{FN}$ stands for the Shakhnovich FN and 
G\={o}$_\mathrm{FN}$ stands for the G\={o} FN. Accordingly, the second column displays the contact's 
mean energy in S$_\mathrm{FN}$; the contact's mean energy in the set of contacts that are common to 
the Shakhnovich nuclei and G\={o}$_\mathrm{FN}$ is shown in the third column. The fourth column 
refers to the set of contacts that are in S$_\mathrm{FN}$ but not in G\={o}$_\mathrm{FN}$ and finally, 
in the fifth column, one considers the contacts that are in G\={o}$_\mathrm{FN}$ but not in S$_\mathrm{FN}$. 
\label{t2}}
\end{table*} 

\subsection{The `topological' role of the folding nucleus}

Despite clear differences, which are driven by contact
stability, the three folding nuclei are nonetheless topologically
similar. The residues that participate in the set of native contacts
forming the folding nuclei split into two groups located in
different regions of the protein chain. Indeed, in all cases
there is a group of 4 residues located in one region of the chain that
comprises residues 2 to 11 and there is another group of 6 (or 8)
residues located in a distant part of the chain that extends between
residue 27 and residue 46. This is illustrated in Figure~\ref{nat}
where the residues whose number along the sequence is less than 12 are
colored in blue while those whose number along the sequence is larger
than 26 are colored in red. It then follows that more than
two thirds of the
contacts that make up the folding nuclei are non-local contacts whose
range lies between 18 to 30 units of backbone separation. In the three
protein models the FN performs the same `topological' role, that of
linking residues located in two distant parts of the protein
chain.

\section{Conclusions}

In the present work we have proposed and discussed in detail a
methodology to the identify the folding nucleus (i.e. a specific subset of
native contacts which, once formed, prompts very rapid and highly
probable folding) in small lattice proteins and applied it to
investigate the nucleation mechanism of three model proteins with
chain length N=48. We have found that a folding nucleus (FN) which is
solely driven by the native fold's topological features (as it happens
in the G\={o} model) is not globally robust with regard to protein
sequence. The latter distinguishes native
contacts, based on the stability of their interaction
energies, and the nucleation pattern is biased towards the most stable
contacts. In other words: in a (more realistic) lattice model,
like a sequence-specific one, the FN is, to some
extent, formed by the most stable contacts, and the presence of
other less stable contacts in the FN is uniquely
determined by the fold's topology. However, we have found that,
independently of protein sequence, the residues forming the three folding nuclei are distributed along the protein chain in a similar and well defined manner. Accordingly, the nucleation mechanism comprises the coalescence of two distinct and distant parts of the protein chain through the establishment of the long range interactions corresponding
to the non-local contacts forming the FN. Therefore we conclude that the fold's topology
determines, to a large extent, the overall position of the FN
in the protein chain. However, as shown by Tiana {\it et al.}~\cite{tiana},
sequences as dissimilar as ours may have a different set of key
residues ({\em e.g.} residues 6 and 35 in our models) in the
FN, which may lead to the latter being topologically distinct. 
\par 
A particularly interesting finding of this work regards the 
existence of 2 residues which, in the three model systems, are
involved in about 30\% of the contacts forming the FN and
appear to be determinant in ensuring fast folding. We
speculate that the network of native contacts formed by these residues
is sufficient to determine the overall fold of the protein in a way
that is similar to that found by Vendruscolo {\it et al.}~\cite{40} for a
98-residue protein model off-lattice.\par    
Previous simulation efforts on lattice models have focused on
smaller (namely N=28~\cite{5} and N=36~\cite{4}) as well as on proteins 
with the same chain length~\cite{39}. We have found that the size of the 
FN is similar to the size of the nuclei identified by
Shakhnovich and collaborators (containing between 8 and 11 native
contacts) which suggests that, at least for small proteins,
the size of the FN does not depend on the size of the chain. This
could provide an explanation for the small correlation between
chain length and folding rates found in real proteins~\cite{41,gal,pra}.\par  
 
Generalizations of the methodology described here, may be useful to
investigate the folding pathways of model proteins. A very preliminary
analysis of our data indicates that there is a higher degree of structural
similarity among the EFCs of the Shakhnovich model than among those of
the G\={o} model. Indeed, we have determined how many different native contacts
exist between each pair of conformations in the three ensembles that were
used to identify the FNs (i.e. in the three ensembles of EFCs) and
computed its mean value over the total number of possible pairs. We have
found that, on average, two EFCs in the G\={o} model differ by 11.3 native
contacts. Sequences 1 and 2, on the other hand, differ by 9.7 and 7.2 native
contacts respectively. We speculate that the higher structural similarity
between conformations in the Shakhnovich model may be related to a smaller
number of rapid folding pathways. However, a definite conclusion requires
further quantitative analysis.

\section{Acknowledgements}

PFNF would like to thank Funda\c c\~ao para a Ci\^encia e Tecnologia
(FCT) for financial support through grant SFRH/BPD/21492/2005. This
work was also supported by FCT's grants POCI/FIS/55592/2004 and POCTI/ISFL/2/618. RDMT wishes
to thank Eugene Shakhnovich and Guido 
Tiana for helpful and elucidating discussions.

\end{document}